# QUARK AND LEPTON MIXING MATRICES: MANIFESTATIONS OF A VIOLATED MIRROR SYMMETRY

*Igor T. Dyatlov*
*Petersburg Institute of Nuclear Physics*
*Scientific Research Center "Kurchatov Institute", Gatchina, Russia*

Existence of heavy mirror analogs of ordinary fermions could explain a thinkable paradox within the Standard Model direct parity violation. This paradox is an objectionable possibility to distinguish physically the left-handed and right-handed coordinate systems. Arguments are presented here that mirror states can also participate in the formation of the observed SM quark and lepton mass spectra and their weak mixing properties. 1. With participation of mirror generations, the quark mixing matrix is similar to the experimentally observed form. The latter is determined by the restrictions imposed by the weak symmetry $SU(2)$ and by the quark mass hierarchy. 2. Under identical conditions and with participation of mirror particles, the lepton mixing (neutrino mixing) can become very different from its quark analog, the Cabibbo-Kobayashi-Maskawa matrix, i.e., it can acquire the qualities prompted by experiment. Such character of mixing evidences in favor of the inverse SM neutrino mass spectrum and the Dirac (non-Majorana) nature of neutrino.

## 1. Introduction

The invention of parity violation in weak interactions gave rise to a paradox which puzzled yet the discoverers of the phenomenon Lee and Yang [1] and was the subject of a number of subsequent studies [2].

Non-conservation of parity means absolute distinction of the left-handed (L) and right-handed (R) coordinate systems—that is, a possibility of distinguishing them physically. For instance, in Wu's experiment [3] with the decay of radioactive $Co^{60}$ in magnetic field, the asymmetry of electron emission with respect to magnetic field allowed field direction to be determined physically: "where more (fewer) electrons fly". The direction of magnetic strength (axial vector) depends upon which coordinate system, L or R, is used to calculate rot $\vec{A}$ ($\vec{A}$ is electromagnetic potential). Determining this direction physically means specifying the coordinate system.

In their second study later in 1956 [1], Lee and Yang proposed a solution to the above paradox which consisted in transposing the geometry, that is, attributing the coordinate system, to the properties of particles themselves. Lee and Yang supplemented the observed system of particles with a system which was identically symmetric but had opposite weak properties. If this symmetry were broken (for example spontaneously), new particles, being very heavy, could not emerge under the conditions in question. They would have emerged in other circumstances, and the particles with weak properties of our world would have been heavy.

___________
e-mail: dyatlov@thd.pnpi.spb.ru

Subsequently, such systems were called mirror systems (see review article [2] and references therein). The significant number of studies devoted to this topic, the large variety of ideas, models, and approaches to this problem [4] indicate that the concepts existing to date with respect to the direct non-conservation of parity do not satisfy many physicists despite the general recognition and successful approbation of the Standard Model (SM). The recent studies (2013) on mirror states discuss the possibility of observing them at LHC [5].

The objective of a previous paper by the author [6] was to prove the statement that the weak-mixing matrix (WMM)—Cabibbo–Kobayashi–Maskawa (CKM) matrix—and its qualitative structure could evidence in favor of the real existence of heavy mirror generations precisely of the type proposed by Lee and Yang for the solution of parity paradoxes.

In this article, it is shown that the observed properties of the lepton WMM (neutrino mixing angles [7]), which are very dissimilar to the CKM matrix, can be interpreted by a mechanism in which heavy mirror generations of leptons play a crucial role. Moreover, the mirror mechanism leads to concrete conclusions about the properties of neutrinos themselves, namely, their Dirac (non-Majorana) nature and the so-called "inverse hierarchy" of the neutrino mass spectrum [7, 8]. Without this mechanism, the character of mixing would be different. This mechanism also provides new evidence for the extraordinary smallness of neutrino masses.

Mass hierarchies of quark and charged leptons are of greatest importance. They determine the structure of CKM matrix nondiagonal elements and, at the same time, ensure that the lepton matrix (with participation of Majorana mass terms) acquires an entirely different form.

Many authors tried to determine the relationship between mass hierarchies and the WMM [9] and to devise the dynamics for formation of the hierarchical spectrum (see review articles [10] and references therein). In our study [6], we did not build the unknown origin dynamics but simply assumed a mass hierarchy for charged mirror generations. The purpose was only to investigate the role of such a spectrum in the formation of the SM quark and lepton WMM. It is this role which is of crucial importance.

At that, any qualitative properties of mixing matrices do not result from fitting of constants or additional relations and ratios of the parameters, as is mostly the case in a number of studies [10]. It is only the mass matrix structure built for a scenario involving intermediate mirror particles which is of importance (Fig. 1). The presence of the chiral group $SU_L(2)$, forming the basis of SM weak interactions, is also important.

The definitive role of this very structure is emphasized by the fact that the number of constants participating in the parametrization in question can greatly exceed the minimum number required for the representation of a totally arbitrary mass matrix (see Section 2). Let us note again:

complete qualitative correspondence to the WMM form does not depend upon the number of parameters and their values and complexities.

This fact renders numerical fitting of the parameters in question uninteresting, bringing no clarification. So far, it is ambiguous and always possible. Restrictions must appear from concrete conditions in dynamic models of mirror symmetry breaking.

In Section 2 of this study, we describe the general procedure for possible inclusion of mirror generations in the problem of quark (schematic repetition of [6]) and lepton spectra and mixing matrices. In Section 3, the neutrino mass matrix is considered in relation with the mirror generations of leptons and mirror symmetry breaking. Arguments in favor of the Dirac nature and inverse hierarchy of the SM neutrino spectrum are presented. In Sections 4 and 5, the lepton WMM is calculated first for the neutrino spectrum with a simple inverse mass hierarchy and then for a realistic inverse hierarchy: two heavy, very close levels (degenerate) and one light level, located far away from the first two. In Conclusions (Section 6), rough estimates of constants used in the proposed mechanism are presented.

Appendix contains a brief description of a theoretically interesting case of Majorana neutrinos in the scenario involving mirror particles. It is explained why this case is considered less appropriate for the experimentally observed situation.

This study greatly relies on the results in [6] and therefore contains numerous references to particular formulae from [6]. Although this may cause inconvenience for the reader, we consider reproduction of the rather cumbersome calculations from [6] impractical.

## 2. Inclusion of Heavy Mirror Generations: General Procedure

In [6], it was the interpretation of mass matrix representations and properties at which both mass hierarchy and WMM for quarks were qualitatively reproduced that led to the idea of mirror particles. In this section, we will do the same exercise but in reverse order: we will use the mirror symmetry and its breaking to reproduce mass matrix representations in [6] matching the observed properties.

Let us consider six generations of Dirac fermions, divided into two groups:

$$1. \quad \Psi_{LRa}^{(f)} = \psi_{La}^{(f)} + \Psi_{Ra}^{(f)}, \qquad 2. \quad \Psi_{RLb}^{(f)} = \psi_{Rb}^{(f))} + \Psi_{Lb}^{(f)}, \qquad (1)$$

where $a, b = 1,2,3$ are generation indices, $f = \bar{u}$ (up) and $\bar{d}$ (down) is a flavor, upper and lower families.

Group 1 ($\Psi_{LR}$) particles are doublets of the flavor group $SU(2)$. They vectorially interact with the weak bosons $W_\mu^A$:

$$g(\overline{\psi_L + \Psi_R})\gamma^\mu \tau_A (\psi_L + \Psi_R) W_\mu^A, \tag{2}$$

which is the sum over generations $a = 1,2,3$.

Group 2 ($\Psi_{RL}$) particles, singlets of $SU(2)$, are sterile and do not interact with the weak boson $W$. Components $R$ and $L$ of the same operators (1) are denoted by different letters because the idea of mirror symmetry

$$R \longleftrightarrow L, \quad \psi \longleftrightarrow \Psi, \tag{3}$$

consists in the transposition of the chiral parts $\psi$ and $\Psi$ from the single Dirac operators $\Psi_{LR}$ and $\Psi_{RL}$ to various particles. Prior to symmetry breaking (3), all interactions are considered $R, L$ symmetric. They are determined by the full operators in (1). This is true for all SM interactions with the exception of Yukawa couplings. The latter or their analogs must appear here upon symmetry breaking (3) (see [6]).

The problem posed by the scenario in question is not the development of system dynamics but rather a selection of conditions that would favor the formation of a quark mass matrix necessary for the appearance of the observed structure. Therefore, only the mass terms of the formula

$$\mathcal{L}' = A_a^{(f)} \left( \bar{\Psi}_{LR}^a \Psi_{LRa} \right)^{(f)} + B_b^{(f)} \left( \bar{\Psi}_{RL}^b \Psi_{RLb} \right)^{(f)}, \tag{4}$$

where $A$ and $B$ are particle masses in the mirror symmetry theory, are important for further representation of the mechanism. One can start with a nondiagonal form for Eq. (4) similar to the nondiagonal expression for Yukawa couplings within SM. This does not affect significantly the further scenario.

A different type of mass terms will appear upon mirror symmetry breaking. The simplest example of violation is the presence in the (now) effective Lagrangian of the mass matrix for the states $\Psi = (\Psi_R, \Psi_L)$. This expression violates the $SU(2)$ symmetry and, at the same time, is, generally speaking, nondiagonal in terms of generation indices:

$$\mathcal{L}'' = \left[\mu\left(\bar{\Psi}_L \Psi_R\right)\right]^{(f)} + \left[\mu^+\left(\bar{\Psi}_R \Psi_L\right)\right]^{(f)}. \tag{5}$$

For convenience, let us assume that (5) is reduced to a diagonal form. Then $\mu_n$, $n = 0,1,2$ (as in [6]), are masses of the states $\Psi_n$. Upon diagonalization in (5), the factors $A$ and $B$ in (4) become general matrices determining transitions $\psi_a \leftrightarrow \Psi_n$. From the form in (4) and definitions in (1) we obtain:

$$\mathcal{L}' = \left[A_a^n \bar{\psi}_L^a \Psi_{Rn} + A_n^{+a} \bar{\Psi}_R^n \psi_{La}\right]^{(f)} + \left[B_b^n \bar{\psi}_R^b \Psi_{Ln} + B_n^{+b} \bar{\Psi}_L^n \psi_{Rb}\right]^{(f)}. \tag{6}$$

At large values of $\mu_n$, Eqs.(5) and (6) result in the following representation of the mass matrix for particles $\psi$:

$$m_{LRa}^{(f)b} = \sum_{n=0}^{2} A_a^{(f)n} \frac{1}{\mu_n^{(f)}} B_n^{+(f)b}, \tag{7}$$

since one can neglect the momentum $|p| \approx m \ll \mu$ in the propagator of the intermediate particle $\Psi$ (see Fig. 1). The separable matrix (Eq.7) was the initial form for the investigation in [6]. Particles $\psi$ acquire their masses while passing through intermediate, very massive states $\Psi_n$. We intend to explain the observed hierarchy of masses $m_a$ in terms of the mass hierarchy of $\Psi$ states:

$$\mu_2^{(f)} \gg \mu_1^{(f)} \gg \mu_0^{(f)} \gg |m|. \tag{8}$$

Similar mechanisms of SM fermion mass formation were used in the extended Technicolor [11] and many other works (see [10], as well as the seesaw mechanism [12] or the review article of Maiani [8]). Fig.1 demonstrates the main difference of the mirror mechanism from mechanisms used earlier. With those, transitions of SM particles into heavy fermions did not occur by means of the mass terms $\mathcal{L}'$ of the mirror-symmetric Lagrangian (Fig.1a) but rather through the interaction with specially invented scalar ("Higgs") fields and through their vacuum averages (Fig.1b).

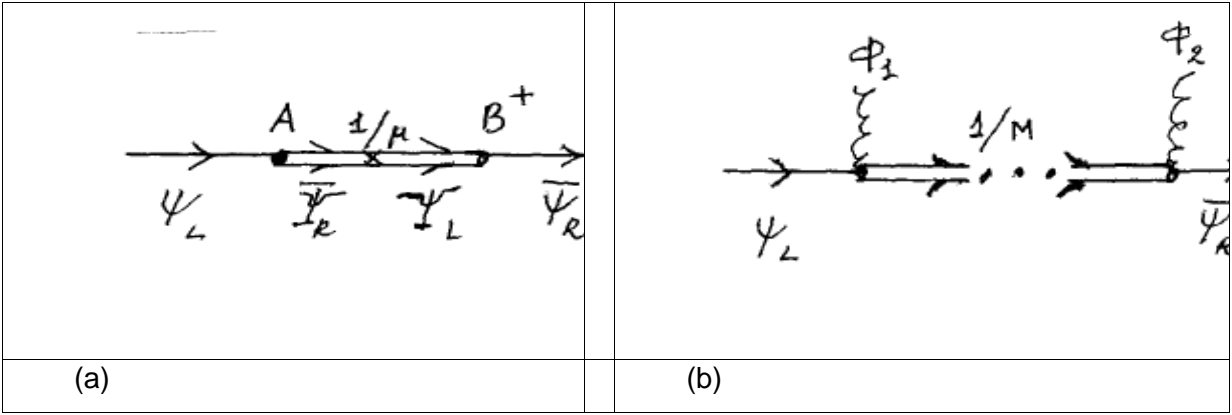

Fig.1: Mechanism of fermion mass formation: in the present article (a) and in [10,11] (b); $\phi_i$ are "Higgs vacuum averages".

Another condition necessary for the qualitative reproduction of the observed WMM from the structure in (7) concerns the nondiagonal form of the matrices $A$ and $B$. These matrices are formed by the vectors $\vec{A}_n$ and $\vec{B}_n$ in the space spanned by the generation indices $a$ and $b$ (in the following, we will omit arrows) and are related with the masses (4) in the mirror symmetry theory. Masses of $A_n^{(f)}$-components of the $SU(2)$ doublets $\Psi_{LR}^{(f)}$ cannot depend upon the flavor $f$ if breaking of the $SU(2)$ invariance takes place only upon appearance of the violating terms in (5). It is imperative that all components of the vectors $A_n^{(f)}$ be independent of the flavor $f$, even in the nondiagonal form in (6), that is, upon breaking of the $SU(2)$-symmetry:

$$A_n^{(\bar{u})} = A_n^{(\bar{d})} \equiv A_n. \qquad (9)$$

This condition is very important. WMM is the scalar product of eigenfunctions of the operator $(mm^+)_{LL}, m$ from Eq.(7), for $\bar{u}$ and $\bar{d}$ quarks (see Eqs.(38) or (4) in [6]). Without imposing the restriction of Eq.(9), we will obtain an arbitrary unitary matrix. Condition (9) automatically results in diagonal elements close to unities and in a hierarchy of nondiagonal elements—that is, the CKM matrix properties.

Let us reiterate that one can start with the nondiagonal form in (4) and diagonal expression in (5). Condition (9) means the $SU(2)$ symmetry of the mirror-symmetric Lagrangian.

Finally, the condition (9) is the principal reason for using mirror states in the Fig.1 mechanism. The invariance $A$ with respect to the flavor group $SU(2)$ can only be introduced into the scheme provided both of the chiral components $\Psi_R$ and $\psi_L$ are doublets of the group. The fermions $\psi = (\psi_L, \psi_R)$ and $\Psi = (\Psi_L \Psi_R)$ differ from each other in mass and weak properties. For the light states $\psi$, the flavor group may be identified with the $SU_L(2)$-chiral group of SM weak interactions. The right-handed currents from Eq.(2) can then be associated only with the heavy particles $\Psi$. Heavy $\Psi$ are mirror analogs of $\psi$-particles in SM.

Rather cumbersome calculations in [6] show that the properties (7)–(9) lead to a hierarchy of light $\psi$-fermion masses and, what is particularly impressive, precisely reproduce the main qualitative properties of the CKM matrix at any values and complexities of $A$ and $B$.

Consequently, the principal result of the proposed mechanism is that the breaking of mirror symmetry (3) transforms the symmetric states

$$\Psi_{LR} = \psi_L + \Psi_R, \quad T_W = \frac{1}{2}, \quad \text{and} \quad \Psi_{RL} = \psi_R + \Psi_L, \quad T_W = 0, \quad (10)$$

(where $T_W$ is a weak isospin) into new cross combinations with opposite and now chiral $SU(2)$ properties – mirror analogs of each other:

$$\begin{aligned} \text{heavy} \quad & \Psi = \Psi_R + \Psi_L, & \text{R is weak current,} \\ \text{light} \quad & \psi = \psi_L + \psi_R, & \text{L is SM weak current} \end{aligned} \quad (11)$$

The properties of the processes behind the origin and decay of $\Psi$ quark mirror generations are schematically outlined in [6] (see also section 6).

Parametrization of Eq.(7) expresses mass matrices of the three generations of $\bar{u}$ and $\bar{d}$ quarks in terms of complex vectors $A_n$ and $B_n^{(f)}$ and real masses $\mu_n^{(f)}$. Thus, we have a total of 18 + 2 x 18 + 6 = 60 independent moduli and phases. Six phases referring to $\psi_R^{(f)}$ and three phases $\psi_L^{(f)}$ (same for $\bar{u}$ and $\bar{d}$ states) do not have physical meaning. The common phase of all $\psi_{L,R}^{(f)}$ does not have any influence. Eq.(7) can include 52 free parameters for two mass matrices in which any values of the elements are parametrized by only 28 independent variables. Therefore, as it was mentioned in Introduction, any experimental values can be easily reproduced.

Let us also note that singling out larger numbers related with the mirror particle masses in Eq.(8) enables one to consider $|A_n|$ and $|B_n|$ as quantities of the same value at all $n$ (this was not so in [6]).

Let us now look at leptons. The known mass hierarchy of charged particles $m_e \ll m_\mu \ll m_\tau$ lacks information on absolute masses of neutrinos. The observed differences of masses squared are assigned to arbitrarily denoted types of massive neutrinos [7, 8]:

$$\begin{aligned}\Delta m_{Sun}^2 &\equiv \Delta m_{12}^2 = m_2^2 - m_1^2 \sim 7.5 \cdot 10^{-5} \text{eV}^2; \\ \Delta m_{atm}^2 &\equiv \Delta m_{23}^2 = m_3^2 - m_2^2 \sim 2.45 \cdot 10^{-3} \text{eV}^2.\end{aligned} \quad (12)$$

From Eqs.(12), it can be seen that one of the states is located far from the other two. Usually, it is considered as state 3. The numeric values indicate that the masses are extremely small (the degeneracy of three large masses is now practically impossible). While ratios of masses of $\bar{u}$ and $\bar{d}$ quarks equal several tens at most, masses of charged leptons exceed possible neutrino masses by $10^7$ to $10^{10}$ times.

The position of masses for states 1, 2, and 3 relative to each other is also unknown. There are two possible ways to interpret experimental data [7, 8]. The small difference of $\Delta m_{12}^2$ masses is assumed to be positive by definition. Then, $\Delta m_{23}^2$ can be both positive ("normal hierarchy") and negative ("inverse hierarchy": $m_2^2 \gtrsim m_1^2 \gg m_3^2$). Neither do we know whether SM neutrinos are Dirac or Majorana types.

The structure of the lepton WMM represented by angles between neutrino types 1, 2, and 3 is completely different from the quark case (see Eq.(43) in Section 4). Those angles are:

$$|\sin\theta_{12}| \approx 0.55; \quad |\sin\theta_{23}| \approx 0.70; \quad |\sin\theta_{13}| \approx 0.15; \quad [7,8]. \quad (13)$$

The above values evidence, in Eq.(43), that the diagonal elements are significantly different from unity, which is characteristic of the CKM matrix.

For charged leptons, the scenario involving mirror intermediate states is fully identical to the quark case. Eq.(7) with lepton parameters $A$ and $B$ must appear. The hierarchy of $e$, $\mu$ and $\tau$ masses is postulated by the inverse (according to (7)) mass hierarchy of mirror charged leptons $\mu^{(ch)}$:

$$\mu_2^{(ch)} \gg \mu_1^{(ch)} \gg \mu_0^{(ch)}. \tag{14}$$

It is in precisely this order of $n = 0, 1, 2$ [6] that expression (14) produces the known hierarchy $m_e \ll m_\mu \ll m_\tau$.

For neutrinos, mirror symmetry breaking can add to (4) both Dirac terms of the type in (5) and new Majorana masses. Let us now write out the Majorana contributions to the effective Lagrangian in a diagonal form, meaning by the factors $A$ and $B$ in the lepton analog of (4) some arbitrary matrices constructed from the complex vectors $A_n$ and $B_n$. Here, too, the condition in (9) is fulfilled:

$$A_n^{(ch)} \equiv A_n^{(\nu)}, \tag{15}$$

that is, the condition of flavor $SU(2)$ symmetry conservation for the lepton $SU(2)$-doublet mass operator $\Psi_{LR}$. The Dirac and Majorana components of the violation may have the following form:

$$\sum_{n=0}^{2} \left\{ \mu_n \left( \bar\Psi_L^n \Psi_{Rn} + \bar\Psi_R^n \Psi_{Ln} \right) + \frac{1}{2} M_{Ln} \left( \bar\Psi_L^n C \bar\Psi_L^{Tn} + \Psi_{Ln}^T C \Psi_{Ln} \right) + \right. \\ \left. + \frac{1}{2} M_{Rn} \left( \bar\Psi_R^n C \bar\Psi_R^{Tn} + \Psi_{Rn}^T C \Psi_{Rn} \right) \right\}, \tag{16}$$

where the symbol $\nu$ is omitted in the operators and in $\mu$ and $M$. $C = -C^T$, $CC^+ = 1$ is the matrix of charge conjugation. For simplicity, let us assume that the $\mu$ and $M$ matrices can simultaneously be reduced to a diagonal form. A general form making the task more complicated would not change the essence of further conclusions.

If the Majorana terms were absent, the system would be fully identical to the quark case. Neither the smallness of neutrino masses nor the different character of mixing has any plausible explanation. The inclusion of $M$ strongly supports argumentation in favor of the mirror mechanism, as is discussed in the next section of this article.

Before we proceed to the actual calculations in the following sections, let us note that, generally speaking, any Majorana terms in (16) seem to be acceptable in the mirror scenario. The terms $|M_L| \neq |M_R|$ will create R,L-asymmetry and parity non-conservation directly, independently of mirror symmetry violations. Two cases:

$$M_R = M_L \quad \text{(17a)} \quad \text{and} \quad M_R = -M_L \quad \text{(17b)} \tag{17}$$

preserve R,L-symmetry and parity. For (17a), parity conservation by the Majorana term in expression (16) requires that all fermions of the system have a parity operator of a certain class, $P = i\gamma_0$. Here, both mirror and normal neutrinos are Majorana particles with different masses. This theoretically interesting case is described in Appendix, since it is not considered by us to fit the observed situation; for this case, CKM type mixing would be more natural. A similar picture is observed at $|M_L| \neq |M_R|$.

The case that matches best the existing data on neutrino characteristics is case (17b). Here all fermions, including neutrinos, belong to a more familiar parity class $P = \gamma_0$ and all neutrinos are Dirac ones (if condition (33') is fulfilled). Masses are arranged in inverse hierarchy, their smallness is supported by sufficiently convincing evidence, and WMM is cardinally different from the CKM matrix.

In the two following sections, we will prove the above statements by constructing a mass matrix and WMM for case (17b).

## 3. Mirror Symmetry Breaking and Neutrino Mass Matrix

In the effective Lagrangian created by mirror symmetry breaking, case (17b) is represented by two types of mass terms. The transition combinations in [6] for neutrino representatives, $\Psi \leftrightarrow \psi$, remain in the mirror-symmetric part. Again, the $SU(2)$ invariance makes one assume that, similar to the quark system, the $f = \bar{u}, \bar{d}$-components of the flavor $SU(2)$-doublets $A_n^{(\nu)}$ and $A_n^{(ch)}$ are equal to each other, Eq.(9), and the singlets $B_n^{(\nu)}$ and $B_n^{(ch)}$ can differ from each other:

$$A_{na}^{(\nu)} = A_{na}^{(ch)} \equiv A_{na}, \qquad B_{nb}^{(\nu)} \neq B_{nb}^{(ch)}. \tag{18}$$

Here again: $a, b$ are generation indices, $n$ are numbers of $A, B$ vectors, $n = 0, 1, 2$.

For the part of the mass terms $\nu$ that do not respect the mirror symmetry, we have expression (16) in which the same Majorana parameters have opposite signs: $M_{Ln} = -M_{Rn} \equiv M_n$. The diagonal form can now be written as follows:

$$\mathcal{L}'' = \sum_{n=0}^{2} \left\{ \mu_n^{(\nu)} \left( \bar{\Psi}_{Rn} \Psi_{Ln} + c.c. \right) + \frac{M_n}{2} \left( \bar{\Psi}_{Rn} C \bar{\Psi}_{Rn}^T - \bar{\Psi}_{Ln} C \bar{\Psi}_{Ln}^T + c.c. \right) \right\}. \quad (19)$$

The difference between the Dirac masses $\mu_n^{(\nu)}$ and the charged mirror lepton masses $\mu_n^{(ch)}$ in Eq.(14) seems to be much smaller than a possible difference between $\mu$ and the Majorana parameters $M$. This choice has been traditional since the first attempt to explain the smallness of neutrino masses—by means of the seesaw mechanism (see review articles [12] and references therein]. This choice is also supported by the moderate, compared to leptons, difference between the masses of $\bar{u}$ and $\bar{d}$ quarks. In the mechanism being discussed, such a quark difference results precisely from the difference of the Dirac mirror masses $\mu_n^{(\bar{u})}$ and $\mu_n^{(\bar{d})}$ in (7) and (8). Let us also note that unlike the seesaw cases in [12], the Lagrangian in Eq.(19) preserves parity.

It is convenient to rewrite Eq. (19) in terms of Majorana operators:

$$\tilde{\Psi}_R = \frac{\Psi_R + C\bar{\Psi}_R^T}{\sqrt{2}}, \qquad \tilde{\Psi}_L = \frac{\Psi_L + C\bar{\Psi}_L^T}{\sqrt{2}}. \quad (20)$$

In terms of operators (20), Eq.(19) has forms that are easy for diagonalization (indices $n$ are omitted):

$$\begin{aligned}
\mu\left(\bar{\tilde{\Psi}}_R \tilde{\Psi}_L + \bar{\tilde{\Psi}}_L \tilde{\Psi}_R\right) &= \mu\left(\bar{\Psi}_R \Psi_L + \bar{\Psi}_L \Psi_R\right), \\
M\left(\bar{\tilde{\Psi}}_L \tilde{\Psi}_L\right) - (L \to R) &= \frac{M}{2}\left(\bar{\Psi}_L C \bar{\Psi}_L^T + \Psi_L^T C \Psi_L\right) - (L \to R).
\end{aligned} \quad (21)$$

Eigenvalues $\lambda$ of the mass Lagrangian (19) for mirror neutrinos are calculated in the representations $\tilde{\Psi}_L$ and $\tilde{\Psi}_R$:

$$\begin{vmatrix} M - \lambda & \mu \\ \mu & -M - \lambda \end{vmatrix} = 0, \quad (22)$$

From this, the masses of these particles are:

$$\lambda_\pm = \pm\sqrt{M^2 + \mu^2} \equiv \pm\lambda. \tag{23}$$

Normalized wave functions corresponding to the masses (23) are readily found:

$$\phi_+ = \frac{1}{N}\left(\tilde{\Psi}_R + \frac{\mu}{M+\lambda}\tilde{\Psi}_L\right), \quad \phi_- = \frac{1}{N}\left(\tilde{\Psi}_L - \frac{\mu}{M+\lambda}\tilde{\Psi}_R\right). \tag{24}$$

At $M \gg \mu$, the normalization factor is practically absent:

$$N = \frac{M+\lambda}{\sqrt{(M+\lambda)^2 + \mu^2}} \approx 1. \tag{25}$$

In fact, two Majorana neutrinos with identical (modulo) masses (23) constitute one Dirac four-component spinor $\Phi$ with the mass $\lambda$:

$$\Phi = \frac{\phi_+ + \gamma_5 \phi_-}{\sqrt{2}}, \quad \bar\Phi = \frac{\bar\phi_+ - \bar\phi_-\gamma_5}{\sqrt{2}}, \quad \Phi_c = \frac{\phi_+ - \gamma_5\phi_-}{\sqrt{2}}. \tag{26}$$

From Eqs.(24) and (25), we have:

$$\tilde{\Psi}_R \simeq \phi_+ - \frac{\mu}{M+\lambda}\phi_-, \quad \tilde{\Psi}_L \simeq \phi_- + \frac{\mu}{M+\lambda}\phi_+, \tag{27}$$

From (26), we find:

$$\phi_+ = \frac{\Phi + \Phi_c}{\sqrt{2}}, \quad \phi_- = \gamma_5 \frac{\Phi - \Phi_c}{\sqrt{2}}. \tag{28}$$

The kinetic parts and gauge interactions are directly rewritten in terms of the variables $\Phi$.

The transition terms for the neutrino operators (6)—$\psi \leftrightarrow \Psi$—can also be expressed in terms of the variables $\Phi$. They are:

$$A\bar{\psi}_L\Psi_R = A\bar{\psi}_L\tilde{\Psi}_R\sqrt{2} = A\bar{\psi}_L\left\{\left(1 - \frac{\mu}{M+\lambda}\gamma_5\right)\Phi + \left(1 + \frac{\mu}{M+\lambda}\gamma_5\right)\Phi_c\right\},$$
$$\bar{\Psi}_L\psi_R B^+ = \bar{\tilde{\Psi}}_L\psi_R\sqrt{2}B^+ = \left\{-\bar{\Phi}\left(\gamma_5 - \frac{\mu}{M+\lambda}\right) + \bar{\Phi}_c\left(\gamma_5 + \frac{\mu}{M+\lambda}\right)\right\}\psi_R B^+. \quad (29)$$

Let us replace the propagators $\langle\Phi,\bar{\Phi}\rangle = \langle\Phi_c,\bar{\Phi}_c\rangle$ by the inverse masses of mirror neutrinos, $\lambda^{-1}$. By multiplying the formulae in Eq.(29) by each other we obtain:

$$m\bar{\psi}_L\psi_R = \bar{\psi}_L A\left\{-\left(1 - \frac{\mu}{M+\lambda}\gamma_5\right)\frac{1}{\lambda}\left(\gamma_5 - \frac{\mu}{M+\lambda}\right) + \left(1 + \frac{\mu}{M+\lambda}\gamma_5\right)\frac{1}{\lambda}\left(\gamma_5 + \frac{\mu}{M+\lambda}\right)\right\}B^+\psi \quad (30)$$

from which the mass matrix for the three SM neutrinos is:

$$(m_\nu)_a^b = \sum_{n=0}^{2} A_{na} \frac{4\mu_n^{(\nu)}}{[\lambda_n(M_n + \lambda_n)]^{(\nu)}} B_n^{(\nu)+b}, \quad (31)$$

—that is, a separable matrix, similar to (7), which has an intermediate state with the effective mass $\sim M^2/\mu \gg M$. In Eq.(31), we reproduce all the earlier omitted indices.

Let us now propose a scenario where neutrinos $\psi$ with the mass matrix (31) are of genuine Dirac nature. For this purpose, let us construct a possible Majorana mass of the $\psi_R, \psi_L$ states using the transition formulae in (29):

$$B\bar{\psi}_R < \left\{\left(\gamma_5 + \frac{\mu}{M+\lambda}\right)\Phi - \left(\gamma_5 - \frac{\mu}{M+\lambda}\right)\Phi_c\right\} C \times$$
$$\times \left\{\Phi^T\left(\gamma_5^T + \frac{\mu}{M+\lambda}\right) - \Phi_c^T\left(\gamma_5^T - \frac{\mu}{M+\lambda}\right)\right\} > \bar{\psi}_R^T B^T, \quad (32)$$

because:

$$\langle\Phi,\Phi_c^T\rangle = \langle\Phi,\bar{\Phi}\rangle C^T = \frac{1}{\lambda}C^T = -\frac{1}{\lambda}C,$$
$$\langle\Phi_c,\Phi^T\rangle = C\langle\bar{\Phi}^T,\Phi^T\rangle = -\frac{1}{\lambda}C. \quad (33)$$

This expression is not equal to zero. For the term $\bar{\psi}_L C \bar{\psi}_L^T$, however, we obtain the same result but with the opposite sign and $B \rightarrow A$ substitution. Then the system with

$$B^v = A \qquad (33')$$

will be the condition for the SM neutrino being Dirac, where its entire mass is defined by the Dirac form (31). At that, generally speaking, $B^v \neq B^{(l)}$.

The contributions $\bar{\psi}_R C \bar{\psi}_R^T$ and $\bar{\psi}_L C \bar{\psi}_L^T$ are cancelled, due to (23), if $\psi_R$ and $\psi_L$ are the chiral parts of the single Dirac particle $\psi = \psi_R + \psi_L$. Then

$$A\bar{\psi}_R \Psi_L + A\bar{\psi}_L \Psi_R = A\bar{\psi}\,(\Psi_L + \Psi_R).$$

It is exactly the Dirac nature of $\nu$ that ensures that the mirror model has the qualitative properties of the observed phenomenology of SM. None of the Majorana scenarios result in the qualitative properties matching the observed picture (see Appendix).

In all formulae, we retain the term $B^v$, as in Eq.(31), but imply the equality (33').

Consequently, despite the presence of Majorana mass contributions in (19), both the mirror $\Phi$ and the SM neutrinos $\psi$ appear here to be Dirac particles.

Let us consider the character of the neutrino spectrum arising from Eq.(31). The Dirac part of the mirror particle mass $\mu_n^{(ch)}$ determines, in the scheme in question, masses of charged leptons. For $\mu_n^{(ch)}$, we have the hierarchy in (14), which is inverse, as per (7), to the hierarchy of leptons. A similar situation is observed for the Dirac quark masses in (8). In all these cases, the smallest mass of a particle is matched by the largest Dirac mass of the mirror state.

This allows us to suggest that the Dirac part of the mirror neutrino mass $\mu_n^{(v)}$ obeys the hierarchy of mirror charged particles in (14), especially considering that the $\bar{u}$ and $\bar{d}$ quark hierarchies are also similar to each other.

In Eq.(31), however, the $\mu_n^{(v)}$ appears in the numerators of the factors of the intermediate states $n$. This means that there is a possibility that the spectrum of the SM neutrinos is inverse: the heaviest of the $\mu_n^{(v)}$ masses corresponds to the heaviest of the $m_n^{(v)}$ masses. Of course, this possibility is only a suggestion based on the assumption that the other numerous factors, $A, B$ and $M$, will not change the tendency of $\mu$. Meanwhile, it is precisely the inverse hierarchy that is prompted by the formulae and is, as we will see later, absolutely preferable in the mirror scenario since it qualitatively corresponds well to the observed neutrino WMM properties.

In conclusion, let us note the presence of the large mass squared $\sim M^2$ in the denominator in Eq.(31). Owing to it, the extraordinary smallness of SM neutrino masses (from Eq.(31), $m_\nu \sim m_{ch}(\mu/M)^2$), does not require looking for enormous fits for $M$ in (17b). For Dirac neutrinos of Eqs.(19) and (31), the smallness of $\psi$ masses is associated with the square of the large mass rather than with its enormous value.

The reader is referred to Conclusions for a discussion of possible numbers and scales.

## 4. Mirror Symmetry Breaking and Neutrino Mixing

The realistic inverse spectrum of SM neutrino masses includes two closely located high levels and one lower level, located far away from the first two (Fig.2a).

$$m_{\nu_1} \approx m_{\nu_2} \gg m_{\nu_3}. \tag{34}$$

Before proceeding to the analysis of this complicated case, let us consider a simpler, in terms of calculation, case of a full inverse hierarchy for all three levels (Fig.2b).

$$m_{\nu_1} \gg m_{\nu_2} \gg m_{\nu_3}. \tag{35}$$

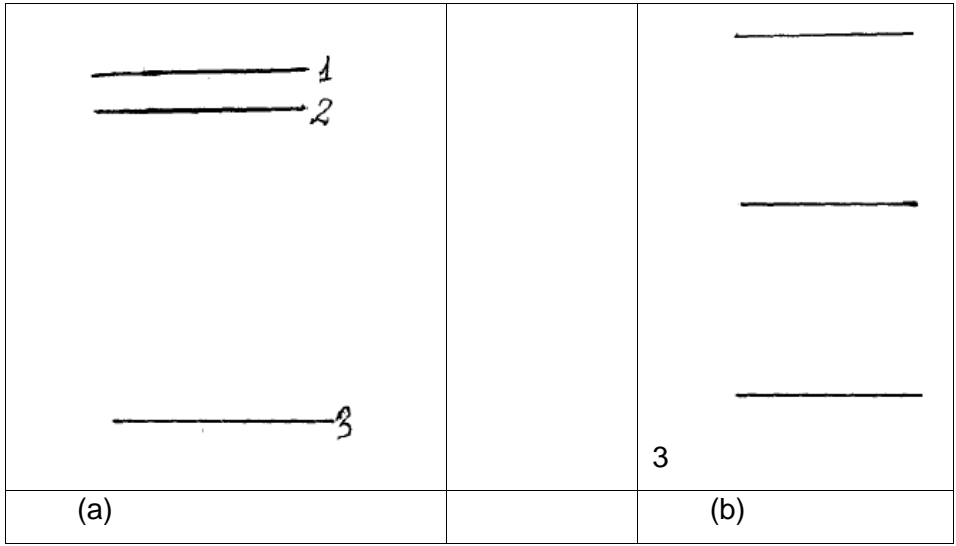

Fig.2: Inverse hierarchies of neutrino states: Realistic spectrum in Section 5 (a); model spectrum in Section 4 (b).

Case (35) preserves all the features of the formation mechanism for a mixing matrix, dissimilar to the CKM matrix, and makes possible the transition to 1-2 degeneracy (34).

Spectrum (35) allows direct use of the formulae from [6], with minimum modifications. Let us denote the $B_n$ vectors from [6] as $B'_n$ to avoid confusion with the $B_n$ vectors in the present paper. Their relationship is:

$$B'_n = \frac{4\mu_n}{\lambda_n(M_n + \lambda_n)} B_n \approx \frac{2\mu_n}{M_n^2} B_n. \qquad (36)$$

The relation between $B'_n$ with different $n$ corresponds to the $\mu_n$ hierarchy in (8) and (14):

$$B'_0 \ll B'_1 \ll B'_2, \qquad (37)$$

and is inverse to the conditions used in [6]. Therefore, the formula for the inverse hierarchy in (35) is obtained from the formulae in [6] by simple transposition of the indices $0 \leftrightarrow 2$. The mixing matrix:

$$U_{MN} = \left(\bar{\Phi}_M^{(\bar{u})}, \Phi_N^{(\bar{d})}\right), \quad M = e, \mu, \tau, \quad N = \nu_1, \nu_2, \nu_3, \qquad (38)$$

is determined by scalar product of the left-hand eigenfunctions of mass matrix squares $(mm^+)_{LL}$ respectively for the $\bar{u}$ and $\bar{d}$ families. The functions $\Phi_M^{(f)a}$ are also vectors in the space spanned by the generation indices $a = 1, 2, 3$. In [6], it is shown (see also Eqs.(29) to (31)) that in the lowest order of the hierarchy (36), the left-hand functions of the separable operators (7) or (31) are determined only by the vectors $A$ that do not depend on the flavor $f = \bar{u}, \bar{d}$ Eq.(18).

In charged leptons with the normal hierarchy of mirror state masses (14), normalized wave functions of SM particles coincide directly with the major contributions from the formulae in [6] (see Eqs.(28) to (30) in [6]).

$$\Phi_\tau = \frac{A_0}{\sqrt{|A_0|^2}}; \quad \Phi_\mu = \frac{|A_0|}{|D_2(|A_0|^2,|A_1|^2)|^{1/2}}\left\{A_1 - \frac{(A_0^+, A_1)}{|A_0|^2}A_0\right\};$$
$$\Phi_e = \frac{1}{|D_2(|A_0|^2,|A_1|^2)|^{1/2}}[A_0^+, A_1^+]. \tag{39}$$

Here, we used the regular denotations for the scalar and vector products of the complex vectors $A_n$. The function $D_2(|A_0|^2, |A_1|^2)$ is the determinant:

$$D_2(|A_0|^2, |A_1|^2) = \begin{vmatrix} (A_0^+, A_0) & (A_0^+, A_1) \\ (A_1^+, A_0) & (A_1^+, A_1) \end{vmatrix}.$$

The orthonormality of the $\Phi$ functions and their equality at $A^{(\bar{u})} = A^{(\bar{d})}$, Eq.(9), led, in [6], to the appearance of unities in the diagonal elements of the WMM and, as a consequence, the hierarchy of nondiagonal terms.

For the inverse hierarchy (37), the eigenfunctions of the operator (31) for neutrino states are obtained from Eq.(39) by transposition of the indices 0 and 2:

$$\Phi_{\nu_1} = \frac{A_2}{|A_2|}; \quad \Phi_{\nu_2} = \frac{|A_2|}{|D_2(|A_2|^2,|A_1|^2)|^{\frac{1}{2}}}\left\{A_1 - \frac{(A_2^+, A_1)}{|A_2|^2}A_2\right\};$$
$$\Phi_{\nu_3} = \frac{1}{|D_2(|A_2|^2,|A_1|^2)|^{1/2}}[A_2^+, A_1^+]. \tag{40}$$

The elements of the unitary matrix (38) are readily determined by the moduli and scalar products of the vectors $A_0, A_1, A_2$. We can reduce the complexities to one phase, by singling out the observable phases, and find, by fitting, three independent real-valued combinations of parameters, as should be the case for the WMM of Dirac particles.

One can obtain a simple and representative version of this procedure by neglecting the complexities (i.e., CP violation) and restricting oneself to the consideration of the real vectors $A_n$. In this case, we see from Eqs.(38) to (40) that in the lowest order of the mass hierarchy, the WMM does not depend on the length of the vectors, which means that it is independent of particle masses and mass ratios (compare with Eqs. (23) - (25), and (36), (37) in [6]). The WMM is dependent only on the angles $\alpha_{01}, \alpha_{02}, \alpha_{12}$ between the vectors $A_0, A_2, A_2$. It is only these angles that parametrize the matrix $U$. In the lowest approximation, not only the diagonal but all of the elements are non-zero. This is the main difference of lepton mixing from the CKM matrix where in this approximation only the diagonal elements appear equal to unity.

Calculations lead to the following structure ($c_{01} = \cos\alpha_{01}$, $s_{01} = \sin\alpha_{01}$, and so on):

|   | $\nu_1$ | $\nu_2$ | $\nu_3$ |   |
|---|---|---|---|---|
| $e$ | $c_{01,2}$ | $c_{01,2}\frac{c_{12}}{|s_{12}|}$ | $c_{01,12}$ | (41) |
| $\mu$ | $c_{01,02}\|s_{02}\|$ | $\frac{1-c_{12}^2-c_{01}^2+c_{01}c_{02}c_{12}}{|s_{01}\cdot s_{12}|}$ | $c_{12,0}\frac{c_{01}}{|s_{01}|}$ |   |
| $\tau$ | $c_{02}$ | $c_{02,12}\|s_{02}\|$ | $c_{12,0}$ |   |

where $c_{01,2}$ is the cosine of the angle between the perpendicular to the $A_0, A_1$ plane, and the vector $A_2$; $c_{01,12}$ is the cosine of the angle between the perpendiculars to the $A_0, A_1$ and $A_1, A_2$ planes. These functions can be expressed in terms of the angles $\alpha_{01}, \alpha_{02}$ and $\alpha_{12}$:

$$s_{0,12}^2 = \frac{c_{01}^2 + c_{02}^2 - 2c_{01}c_{02}c_{12}}{s_{12}^2} ; \quad c_{01,12} = \frac{c_{02} - c_{01}c_{12}}{|s_{01}\cdot s_{12}|}. \qquad (42)$$

Other analogous functions from the matrix (41) are obtained by transposition of indices. Formulae (42) allow us to check the unitarity of (41). One should note, however, that use of perpendiculars to planes for determining the elements requires that signs be specified with scrutiny. In Eq.(41), we have in fact the moduli of elements written out. One can easily check only the sums of squares of row element moduli.

Formula (41) can be compared to the matrix $U$ in the commonly used angle parametrization (13), see [7,8]. Assuming that here, too, the CP-violating phase is equal to zero, we have (now $c_{12} = \cos\theta_{12}$, $s_{12} = \sin\theta_{12}$ and so on):

$$\begin{vmatrix} c_{12}c_{13} & s_{12}c_{13} & s_{13} \\ -s_{12}c_{23}-c_{12}c_{23}s_{13} & c_{12}c_{23}-s_{12}s_{13}s_{23} & s_{23}c_{13} \\ s_{12}s_{23}-c_{12}c_{23}s_{13} & -c_{12}s_{23}-s_{12}s_{13}c_{23} & c_{23}c_{13} \end{vmatrix}. \tag{43}$$

Of course, both parametrizations—(41) and (43)—can equally be used, since either one contains three independent parameters. In (43), there is no physics. One can find the relative location of the $A_1$, $A_2$ and $A_0$ vectors based on the experimental data in (13), however, this knowledge is futile unless used in some concrete mirror symmetry violation model. For instance, the smallness of the $\theta_{13}$ angle [8] means that the $A_0, A_1$ and $A_1, A_2$ planes are approximately perpendicular to each other. Note that this element, as well as the entire last column in (41), does not change in the case of the realistic spectrum in Fig.2a.

## 5. Lepton WMM Calculation for Neutrino Mass Degeneracy Case, Fig. 2a

The calculations in this section are even more intimately related with the results in [6]. Let us repeat again: to use the formulae for the neutrino mass inverse hierarchy case, Fig.2, one should transpose the $n$ indices 0 and 2 in the formulae in [6].

Let us determine the conditions imposed on the neutrino parameters $A_n$ and $B'_n$ by the spectrum in Fig.2a. Of the three eigenvalues $\rho_N$ of the Hermitian matrix $(mm^+)_{LL}$, in which $m$ is specified by (31), the least of the characteristic equation roots $\rho_{III} = m_{\nu_3}^2$ can, in the lowest hierarchy order, be taken as zero. As follows from the formulae for the eigenvalues (Eqs.(23) - (25) in [6]), this means that $B'_0 \simeq 0$ (inverse hierarchy!). Consequently, $\text{Det}(mm^+)_{LL} = 0$, and the third-order characteristic equation is reduced to quadratic. Its roots are:

$$\rho_{I,II} = \frac{1}{2}\text{Tr}(mm^+)_{LL} \pm \sqrt{\frac{1}{4}[\text{Tr}(mm^+)_{LL}]^2 - \Sigma}, \tag{44}$$

where $\Sigma$ is a sum of the principal minors $(mm^+)_{LL}$.

In the expressions for the trace, Eq.(19) in [6], and for the sum of the principal minors of Eq.(20) in [6], we should assume $B'_0 = 0$ (i.e., $B_2 = 0$ in [6]). Then we obtain:

$$\mathrm{Tr}(mm^+)_{LL} = |A_2|^2|B_2'|^2 + 2\mathrm{Re}(A_2, A_1^+)(B_2'^+, B_1') + |A_1|^2|B_1'|^2$$
$$\Sigma = \left(|A_1|^2|A_2|^2 - |(A_1^+, A_2)|^2\right)\left(|B_1'|^2|B_2'|^2 - |(B_1'^+, B_2')|^2\right). \tag{45}$$

Similar to Section 4, let us consider the vectors $A_n$ and $B_n'$ real-valued. Then the radical expression under the root in (44) is reduced to the form:

$$\frac{1}{4}\left(A_2^2 B_2'^2 - A_1^2 B_1'^2\right)^2 + A_2^2 A_1^2 (B_2', B_1')^2 + B_2'^2 B_1'^2 (A_2, A_1)^2 + \\ + A_2^2 B_2'^2 (A_2, A_1)(B_2', B_1') + A_1^2 B_1'^2 (A_2, A_1)(B_2', B_1'). \tag{46}$$

Expression (46) can be equal to zero only at:

$$|A_1| = |A_2|, \quad |B_1'| = |B_2'|, \quad \cos\alpha_{12} = -\cos\beta_{12}, \tag{47}$$

where $\beta$ are angles between the vectors $B_n'$. The equality $|A_1| = |B_2'|$, and so on, contradicts the $SU(2)$ properties of $A$ and $B$.

Let us now proceed directly to the problem of degeneracy removal in Fig.2a. According to the standard quantum-mechanical procedure [13], the equation

$$\left[(mm^+)_{LL\,a}^{(0)\,b} + (mm^+)'_{LL\,a}^{\,b}\right]\Phi_b^{(I,II)} = (\lambda_0 + \lambda'_{I,II})\Phi_a^{(I,II)} \tag{48}$$

is solved by perturbation theory relative to the negligible correction $(mm^+)'_{LL}$. For this purpose, one should choose two solutions, orthogonal to each other, of zero approximation—$\phi_1$ and $\phi_2$. Then we can find correct functions of zero approximation for degenerate states I and II as follows:

$$\Phi^{(I)} = \alpha_I \phi_1 + \beta_I \phi_2, \quad \alpha^2 + \beta^2 = 1, \\ \Phi^{(II)} = \alpha_{II} \phi_1 + \beta_{II} \phi_2. \tag{49}$$

The "energies" $\lambda'_{I,II}$ are found from the characteristic equation:

$$\mathrm{Det}\{(\phi_i, (mm^+)'_{LL}\phi_k) - \lambda'\delta_{ik}\} = 0, \tag{50}$$

whereas the relation of the coefficients in (49), $\alpha$ and $\beta$, is expressed by the formula:

$$\alpha = \frac{(\phi_1, (mm^+)'_{LL}\phi_2)}{(\phi_1, (mm^+)'_{LL}\phi_1) - \lambda'}\beta = \frac{\lambda' - (\phi_2, (mm^+)'_{LL}\phi_2)}{(\phi_2, (mm^+)'_{LL}\phi_1)}\beta. \tag{51}$$

In the task being discussed, expressions for $(mm^+)_{LL}^{(0)}$ and $(mm^+)'_{LL}$ are found from Eq.(27) in [6]. Using (47) and denoting the vectors in terms of unit vectors $n$, $n'$: $A_a = An_a$, $B_a = B'n'_b$, we obtain:

$$\begin{aligned}
\left[(mm^+)_{LL}^0\right]_{ab} &= A_2^2 B_2'^2\{n_{2a}[n_{2b} + (n'_1, n'_2)n_{1b}] + n_{1a}[n_{1b} + (n'_1, n'_2)n_{2b}]\}, \\
\left[(mm^+)'_{LL}\right]_{ab} &= A_2 B_2' A_0 B_0'\{[n_{2a}(n'_0 n'_2) + n_{1a}(n'_0 n'_1)]n_{0b} + \\
&\quad + n_{0a}\left[(n'_0 n'_2)n_{2b} + (n'_0 n'_1)n_{1b}\right]\}.
\end{aligned} \tag{52}$$

Note that according to Eq.(47), $(n_1, n_2) = -(n'_1, n'_2)$, $|B'_0| \ll |B'_2| = |B'_1|$.

As $\phi_1$ and $\phi_2$, we can choose states (40); their expressions in terms of unit vectors $n$ are:

$$\phi_1 = n_2, \quad \phi_2 = \frac{n_1 - (n_1, n_2)n_2}{|\sin\alpha_{12}|}, \quad |\sin\alpha_{12}| \equiv |[n_1, n_2]|. \tag{53}$$

The eigenfunction of non-degenerate state 3 remains the same as that in the hierarchy of Fig.2b; that is:

$$\phi_3 = \Phi^{(III)} = \frac{[n_1, n_2]}{|\sin\alpha_{12}|}. \tag{54}$$

It is easy to check that $\phi_1$ and $\phi_2$ for the condition in (47) are orthonormalized eigenfunctions of the operator $(mm^+)_{LL}^{(0)}$ with a degenerate eigenvalue:

$$\lambda_0 = A_2^2 B_2'^2 \sin^2 \alpha_{12}. \tag{55}$$

The matrix elements $(mm^+)'_{LL}$ required for the calculation of $\lambda'$ and coefficients $\alpha$ and $\beta$ are obtained with the help of Eqs.(52) and (53):

$$\begin{aligned}
\left(\phi_1, (mm^+)'_{LL} \phi_1\right) &= 2A_2 B_2' A_0 B_0' (n_0, n_2) \left[(n_0', n_2') + (n_1, n_2)(n_0', n_1')\right], \\
\left(\phi_2, (mm^+)'_{LL} \phi_2\right) &= 2A_2 B_2' A_0 B_0' (n_0', n_1') \left[(n_0, n_1) - (n_0, n_2)(n_1, n_2)\right], \\
\left(\phi_1, (mm^+)'_{LL} \phi_2\right) &= \left(\phi_2, (mm^+)'_{LL} \phi_1\right) = \frac{A_2 B_2' A_0 B_0'}{|\sin \alpha_{12}|} \Big\{ (n_0, n_1)(n_0', n_2') + \\
&\quad + (n_0, n_2)(n_0', n_1') + (n_1, n_2) \left[(n_0, n_1)(n_0', n_1') - \right. \\
&\quad \left. - (n_0, n_2)(n_0', n_2') - 2(n_0, n_2)(n_0', n_1') \right] \Big\}.
\end{aligned} \tag{56}$$

Eq.(56) permits calculation of all quantities involved in the problem of level I and II degeneracy. With the functions $\Phi^{(I)}$ and $\Phi^{(II)}$ of Eq.(49) found, let us also determine the mixing matrix in (38). It is clear that no qualitative changes, compared to the matrix (41), occur. In fact, since the function $\Phi^{(III)}$ in Eq.(54) does not change, the last column of the matrix (41), as mentioned earlier, remains the same. We will not provide here the calculations for the two other columns, which, while having no qualitative effect, are very cumbersome.

## 6. Conclusions

Based on the analysis in this article, we can make the following conclusions.

1. Even in the lowest-order approximation in a mass hierarchy, all elements of the lepton WMM are, generally speaking, different from zero and do not depend on the masses or their ratios. This makes lepton mixing completely dissimilar to the CKM matrix.

2. The large effective mass $\sim M^2/\mu \gg M$ of the mirror intermediate state in (31) presents an additional theoretical argument (other than the seesaw mechanism [12]) for explaining the smallness of neutrino masses. The new mechanism preserves parity.
3. This situation is only possible for Dirac neutrinos that have inverse hierarchy. Therefore, it seems to be most appropriate for describing SM neutrinos. It is precisely this case that is prompted by the formulae of the mirror scenario.

The matters of creation and decay of mirror particles, briefly described earlier in [6], require, of course, a deeper and more thorough analysis.

We will only note here that the seemingly most available processes involving mirror particles can occur, in the first place, through the weak interaction (2). This is possible owing to the presence of the small terms, at $m \ll \mu$, in the mixing of mirror particles and SM fermions (see Eqs.(46) in [6]). Such mixing is similar to the small additions in wave functions in Eqs. (24) and (27) in Section 3. In this article, we ignored these additions and termed the quantities $m$, obtained by diagonalization of formulae (7) or (31), as well as $\mu$ and $M$ specified in these formulae, as the masses of the corresponding states, both SM and mirror. For $m \ll \mu \ll M$, this is essentially the case.

At present, no reliable method exists for evaluating the mass of the lightest mirror particle. In our scenario, what is beyond doubt is only that this particle is analogous to the $t$-quark and would appear in processes involving it. Mass evaluation must include an investigation of the effect the value of this mass may have on the cross-section of rare processes (such as the $\mu \leftrightarrow e\gamma$ for leptons), for which very stringent experimental restrictions exist (see [7, 14]). There is also the possibility of additional contributions to the processes producing the Higgs boson.

Let us now evaluate orders of magnitude for all the parameters used in this article, proceeding from the fact that the condition imposed on the mirror masses $\mu \gg m$ requires that $\mu \gg m_t$, which is the mass of the heaviest $t$-quark. Under this condition, the value $\mu \sim 1$ TeV is quite acceptable. For simplicity, let us use this value. Then, we can assume that the parameters for quarks and leptons will have the following ranges:

- Mirror-symmetrical masses $A$ and $B \sim 10^2 - 10^3$ GeV but less than $10^3$ and larger than $m_t$

- Mirror quark and lepton masses, for various "up" generations, $\mu_q^{(u)}, \mu^{(ch)} \sim (10^3 - 10^{7-8})$ GeV
- Mirror quark masses, for "down" generations, $\mu_q^{(d)} \sim (10^4 - 10^7)$ GeV
- Majorana masses $M \sim (10^9 - 10^{12})$ GeV

The author is grateful to Ya. I. Azimov, G.S. Danilov, L.N. Lipatov and M.G. Ryskin for their interest in this work and for very useful discussions. This work was funded by grant RSF No. 14-92-00281.

## Appendix

Case (17a): the equality of the Majorana parameters $M_R = M_L$, after the transition to the Majorana operators in (20), leads to the equation for masses and eigenvalues of mirror particles in the form of the determinant:

$$\begin{vmatrix} M - \lambda & \mu \\ \mu & M - \lambda \end{vmatrix} = 0. \qquad (A.1)$$

Eq. (A.1) determines two different mass values:

$$\lambda_\pm = M \pm \mu. \qquad (A.2)$$

This means that the four components $\tilde{\Psi}_L, \tilde{\Psi}_R$ in Eq.(20) split into two different two-component Majorana particles.

Unlike the functions in (24), the eigenvalues of mirror particles are effectively (at $M \gg \mu$) expressed in terms of the full four-component operators $\Psi = (\Psi_R, \Psi_L)$:

$$\phi_+ \simeq \frac{\tilde{\Psi}_L + \tilde{\Psi}_R}{\sqrt{2}} = \frac{1}{2}(\Psi + C\bar{\Psi}^T); \quad \phi_- \simeq \frac{\tilde{\Psi}_L - \tilde{\Psi}_R}{\sqrt{2}} = \frac{\gamma_5}{2}(\Psi - C\bar{\Psi}^T). \qquad (A.3)$$

The Majorana nature of $\phi_\pm$ is evident (it is convenient to introduce the $i$ multiplier in $\phi_-$).

The particle $\psi$ acquires both the Dirac and Majorana parts of the mass matrix in the process of the transition into the intermediate states $\phi_+$ and $\phi_-$ by means of Eq.(6). The Dirac part of the mass $\psi$ appears as a result of the following transformation (Fig.1):

$$\psi_L \to \bar{\Psi}_R \to \sqrt{2}\,\bar{\tilde{\Psi}}_R \to \langle \phi_+, \bar{\phi}_+ \rangle - \langle \phi_-, \bar{\phi}_- \rangle \to \tilde{\Psi}_L \sqrt{2} \to \Psi_L \to \bar{\psi}_R, \quad (A.4)$$

from which the Dirac part of the mass terms of light (at large $M$) neutrinos is equal to:

$$\sum_n A \left( \frac{1}{\lambda_+} - \frac{1}{\lambda_-} \right) B^+ \;\Rightarrow\; -\sum_n A \frac{\mu}{M^2} B^+. \quad (A.5)$$

The Majorana part results from:

$$\psi_L \to \bar{\Psi}_R \to \sqrt{2}\bar{\tilde{\Psi}}_R \to \langle \phi_+, \phi_+^T \rangle + \langle \phi_-, \phi_-^T \rangle \leftarrow \sqrt{2}\tilde{\bar{\Psi}}_R^T \leftarrow \bar{\Psi}_R^T \leftarrow \psi_L^T. \quad (A.6)$$

For the Majorana operators $\phi$, we have $\phi^T = C_\phi^+$. Then:

$$\langle \phi, \phi^T \rangle = -\frac{1}{\lambda} C. \quad (A.7)$$

The Majorana part of a neutrino mass can symbolically be written in the form:

$$\mathcal{M} \sim \sum_n A \left( \frac{1}{\lambda_+} + \frac{1}{\lambda_-} \right) A \;\Rightarrow\; \sum_n A \frac{2}{M} A, \quad (A.8)$$

plus analogous BB and conjugate terms.

Constructing a matrix equation similar to Eq.(A.1) for the masses of states $\psi$, we obtain the following expressions for the masses $m_\nu$:

$$m_\nu \sim O\left(\frac{1}{M}\right) \pm O\left(\frac{\mu}{M^2}\right), \tag{A.9}$$

Thus, in Eq.(17a) we deal with Majorana particles, both heavy and light. Masses of light particles in (A.9) are in general inversely proportional to masses of heavy mirror neutrinos. We observe a full analogy with the mechanism of charged quark and lepton mass formation. CKM mixing matrix appears to be more appropriate in this case. One sees no specific reason here for the formation of the inverse spectrum of light neutrinos. The hierarchy of mirror Majorana masses is uncertain, and the masses must be very large. At $A$ and $B$ being of the orders of magnitude discussed in Section 6, the $M$ masses are already close to the Planck mass.